\documentstyle[epsfig]{mn}

\def\Msun{\ifmmode{~{\rm M}_\odot}\else${\rm M}_\odot$\fi}
\def\kms{\ifmmode{$~km\thinspace s$^{-1}}\else km\thinspace s$^{-1}$\fi}
\def\Teff{T_{\mathrm eff}}

\def\ee{\end{equation}}
\def\be{\begin{equation}}

\title{The colours of the Sun}

\author[Johan Holmberg, Chris Flynn and Laura Portinari] {Johan Holmberg$^{1}$,
Chris Flynn$^{1}$ and Laura Portinari$^{1}$\\ $^1$Tuorla Observatory,
V\"ais\"al\"antie 20, FI-21500 Piikki\"o, Finland}

\date{\today}

\begin{document}

\maketitle

\voffset=-1.0cm     

\begin{abstract} 
We compile a sample of Sun-like stars with accurate effective temperatures,
metallicities and colours (from the UV to the near-IR). A crucial improvement
is that the effective temperature scale of the stars has recently been
established as both accurate and precise through direct measurement of angular
diameters obtained with stellar interferometers. We fit the colours as a
function of effective temperature and metallicity, and derive colour estimates
for the Sun in the Johnson/Cousins, Tycho, Str\"omgren, 2MASS and SDSS
photometric systems.  For $(B-V)_{\odot}$, we favour the ``red'' colour 0.64
versus the ``blue'' colour 0.62 of other recent papers, but both values are
consistent within the errors; we ascribe the difference to the selection of
Sun-like stars versus interpolation of wider colour-$\Teff$-metallicity
relations.

\end{abstract}


\section{Solar colours from Sun-like stars}

The colours and absolute magnitudes of the Sun in various passbands are a
natural calibration point for a wide range of stellar, Galactic and
extra-Galactic astronomy. The Sun is the closest star, with the best known,
directly measured physical parameters (luminosity, mass, age, radius, chemical
composition). It is the first test-bench for calibrating models of stellar
structure, and all quantities in stellar physics (of special interest here, the
luminosities) are to be scaled to the solar values.  Also when studying more
complex systems, such as external galaxies, one needs the solar magnitudes to
translate the observed integrated magnitudes and colours into physical
luminosities, so as to interpret them with theoretical population synthesis
models.  In an on-going study (Holmberg, Portinari and Flynn, in preparation)
we are determining the surface brightness of the local Galactic disc in bands
from the UV to the near IR, motivated by having a good description of the mass
density of its stellar content (Holmberg \& Flynn 2000, 2004), and we
ultimately aim at estimating the mass--to--light ratio of the local disc, for
comparison to external galaxies, to the Tully--Fisher relation and to models of
the chemical and photometric evolution of disc galaxies.  This study motivated
us to investigate more closely the present status of the solar magnitudes and
colours in the various bands, needed to link between surface brightness and
mass--to--light ratio.

The Sun is such a bright, non-point-like source that measuring its colours in
the passbands adopted for astronomical work is no easy task, as the photometric
devices used for observing other stars cannot be applied to the Sun. One way to
circumvent the problem is to use the solar spectrum (theoretical or observed)
and convolve it with the known filter responses to derive the solar magnitudes
in the various photometric systems. An alternative solution is to search in the
sky for `Sun--like' stars, which have very similar properties to the Sun and
from them infer the solar values.  The study of Sun--like stars was pioneered
by Hardrop (1978), with many more contributions in the 80's\footnote{It may be
worth remarking that the solar colours reported in quite recent textbooks such
as Allen's Astrophysical Quantities 2000, are actually based on references from
the 1980's} and 90's extensively reviewed by Cayrel de Strobel (1996);
following her, we can distinguish:
\begin{itemize}
\item
Sun--like stars: F-G stars with colours ($B-V$, typically) similar to those of
the Sun;
\item
solar analogues : stars with the same Spectral Energy Distribution (SED) of the
Sun, i.e.\ basically stars with the same metallicity and effective temperature
($\Teff$);
\item
solar twins: stars with all the physical parameters (mass, age, chemical
composition, $\Teff$, surface gravity etc.) ``identical" to those of the Sun.
\end{itemize}
The closest ``solar twin'' found so far is HR 6060 (Porto de Mello \& da Silva,
1997), with $(B-V) \sim 0.65$ and $(U-B) \sim 0.18$. Very recently, King,
Boesgaard \& Schuler (2005) have suggested that HIP 78399 ($B-V \sim 0.644$)
could be another candidate.  Solar twins are ideal to define the photometry of
the Sun, but due to their paucity, in practice one resorts to studying solar
analogues more in general.

The most studied by far, yet elusive, of the solar colours is $B-V$: the
reddish values of 0.66--0.69 popular in the 70's and 80's are presently ruled
out, as all recent estimates agree on 0.62--0.65 (Cayrel de Strobel 1996;
Sekiguchi \& Fukugita 2000; Soubiran \& Triaud 2004; Ram\'\i rez \& Mel\'endez
2005b). Within this presently accepted range, studies pre-selecting the solar
analogues, for instance by spectroscopy, favour $(B-V)_{\odot} \sim 0.64-0.65$
(e.g.\ Porto de Mello \& da Silva 1997; Soubiran \& Triaud 2004), while
interpolation within more general $T_{\rm eff}$--colour--[Fe/H] relations for
Sun-like stars tends to yield a bluer $(B-V)_{\odot} \sim 0.62$ (Sekiguchi \&
Fukugita 2000; Ram\'\i rez \& Mel\'endez 2005b). The latter result also implies
that the solar $B-V$ predicted by theoretical model 
atmospheres: $(B-V)_{\odot} \sim 0.65-0.67$ 
(Bessell, Castelli \& Plez 1998; Casagrande et~al.\ 2005, in prep.)
is too red.  Sekiguchi \& Fukugita ascribe
the discrepancy to systematics in the $\Teff$ scale, i.e.\ spectroscopic vs.\
InfraRed Flux Method (IRFM) --- though within their quoted uncertainty
$(B-V)_{\odot} = 0.626 \pm 0.018$, values $\sim$0.64 remain plausible.

We here analyse `Sun-like stars', chosen in a range of effective temperatures
which brackets the effective temperature of the Sun. The colours of the stars,
in a number of passbands, are fitted for dependence on $\Teff$ and metallicity
[Fe/H]. The Sun's colours are then indirectly determined from these relations
for adopted solar values of $T_{\mathrm eff,\odot} = 5777$~K and [Fe/H] $=
0.0$.

Crucial to this procedure is that the effective temperatures of the stars and
the Sun are on the same scale; historically, this has been the basic impasse to
estimating the solar colours indirectly from Sun-like stars. During the last
two years, this situation has fundamentally altered, as interferometric
measurements of stellar diameters have started to become routine at such
instruments as the ESO VLTI (European Southern Observatory's Very Large
Telescope Interferometer), PTI (Palomar Testbed Interferometer) and NPOI (Navy
Prototype Optical Interferometer). These instruments have produced high
accuracy stellar diameters for about 20 main sequence stars from $\approx
1.3$~\Msun\ to $\approx 0.7$~\Msun\ (Kervella et al. 2004).  As a consequence,
the effective temperatures of the stars can be determined in the same manner as
for the Sun, from the physical definition of $\Teff$, in terms of the
bolometric luminosity at the Earth, $f_{\mathrm bol}$, and the angular
diameter, $\theta$, via

\be
f_{\mathrm bol} = \frac{\theta^2}{4} \sigma T_{\mathrm eff}^4
\ee

\noindent where $\sigma$ is the Stefan-Boltzmann constant.

Using 13 stars with directly measured stellar diameters and bolometric fluxes,
Ram\'\i rez \& Mel\'endez (2005a) have determined direct effective
temperatures, $T_{\mathrm eff}^{\mathrm dir}$. They show that the widely used
IRFM temperature scale (Alonso, Arribas \& Mart\'inez-Roger 1996) is a very
good match to $T_{\mathrm eff}^{\mathrm dir}$, but also derive an improved
version of Alonso et al's IRFM temperature scale, showing that the systematic
difference between the IRFM and the directly measured temperature scales is not
significantly different from zero (formally, they obtain a difference of 18 K,
with a dispersion around the mean of 62 K, and an error in the mean of 21 K for
their 10 highest quality stars, in the sense that the IRFM scale is the hotter
one). In Ram\'\i rez and Mel\'endez (2005b), the authors derive a number of
colour-temperature relations for main sequence stars (and giants), from which
estimates of the Sun's colours are presented in a range of photometric
systems. We became aware of this study while undertaking our own, and
eventually made use of their extended set of IRFM temperatures in order to
supplement material we had collected, as described in detail below. Our method
is to fit the colours of Sun-like stars as a function of their $T_{\mathrm
eff}$ and metallicities from Nordstr\"om et al (2004), and then use these fits
to solve for the colours of the Sun, for which we adopt $\Teff = 5777$ K and
[Fe/H] $ = 0$.

\section{Sample of Sun-like stars}

\begin{figure} 
\par\centerline{\psfig{figure=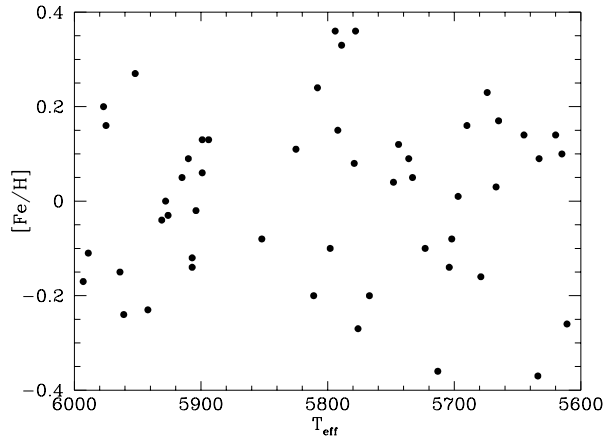,width=6cm,angle=-90}}
\caption{The temperature and metallicity distribution for the sample of
Sun-like stars used in this paper.}
\label{tefeh}
\end{figure}

Stars similar to the Sun have been selected from the sample of Ram\'\i rez and
Mel\'endez (2005a) in the range $5600 < \Teff < 6000$, in order to bracket the
solar effective temperature $\Teff = 5777$~K, resulting in 115 main sequence
stars. For 67 of these, accurate metallicities on a homogeneous system are
taken from Nordstr\"om et al's (2004) study of the kinematics and chemistry of
some 14\,000 nearby F, G and K stars. The photometric metallicities of
Nordstr\"om et al. (2004) have been shown to be in excellent agreement with
spectroscopic ones. When compared to e.g. Edvardsson et al. (1993), the total
dispersion between photometric and spectroscopic [Fe/H] is 0.08 dex.  From this
sample we select stars with a metallicity bracketing the solar one, in the
range $-0.40 <$ [Fe/H] $<$ +0.40 resulting in 52 stars which form our basic
sample of main sequence, Sun-like stars.  Their distribution in $\Teff$ and
[Fe/H] is shown in Figure \ref{tefeh}.

\begin{figure} 
\par\centerline{\psfig{figure=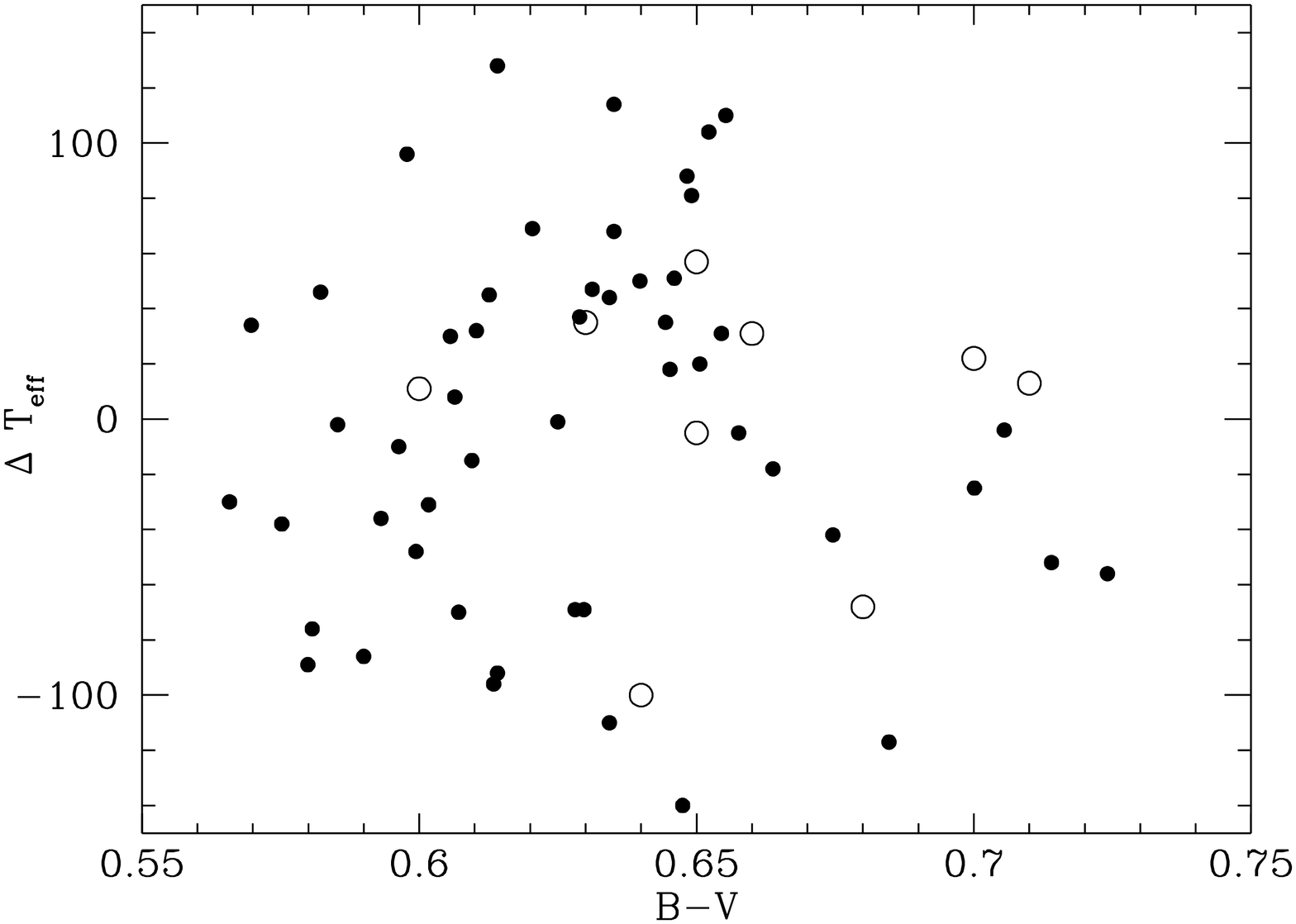,width=6cm,angle=-90}}
\par\centerline{\psfig{figure=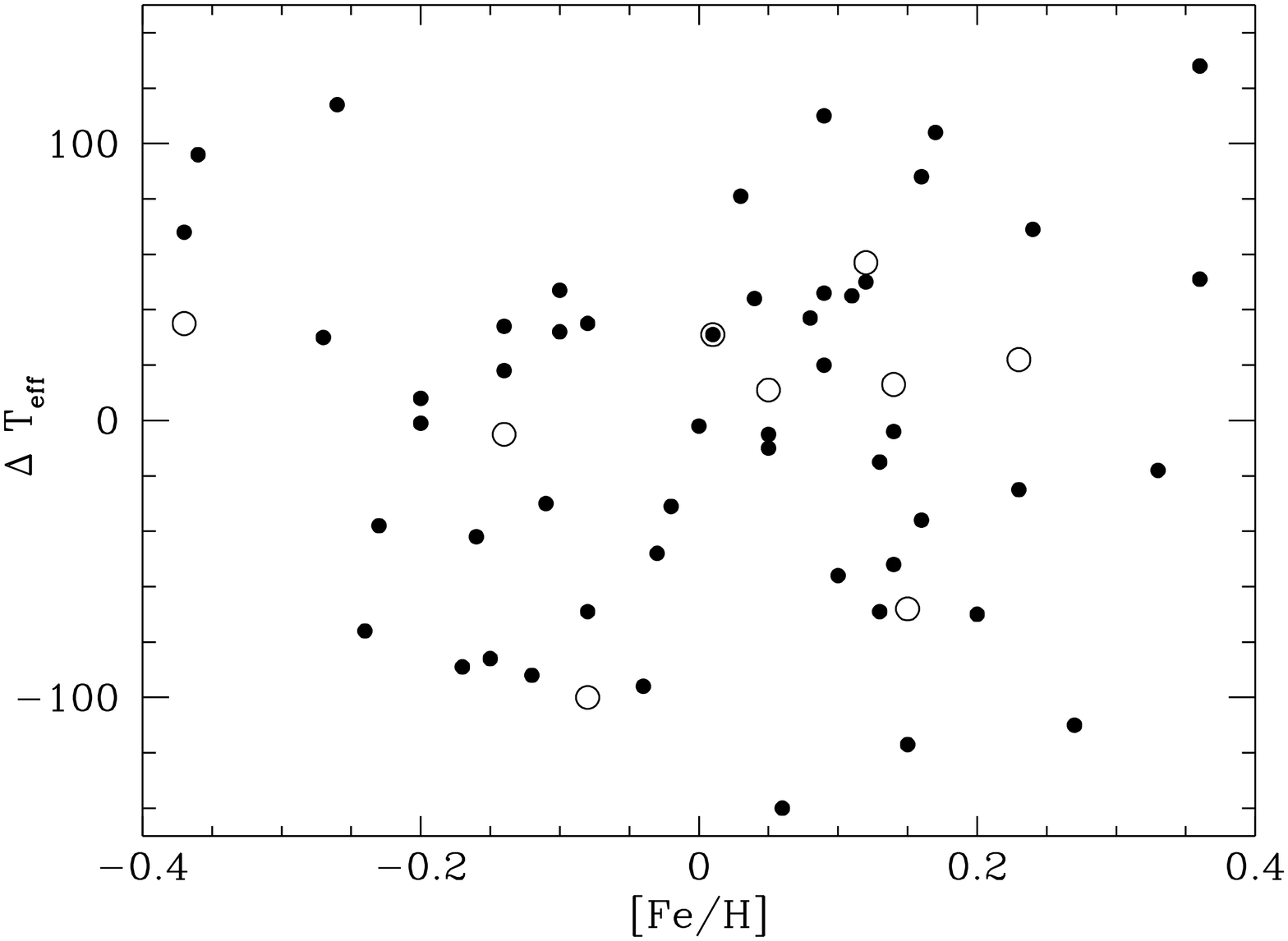,width=6cm,angle=-90}}
\caption{The difference between the IRFM temperatures and the temperatures
fitted as a function of colour and metallicity. Large open symbols show the
direct fit, small symbols the fit using Tycho photometry.}
\label{tefit}
\end{figure}

We collect from the literature wide-band Johnson $UBV$, Cousins $RI$, Two
Micron All Sky Survey (2MASS)
$JHK_s$, Tycho $V_{\rm T}$, $B_{\rm T}$ and medium-band Str\"omgren $ubvy$
photometry for our sample stars, so as to determine indirectly the solar
colours from the ultraviolet to the infrared.  For 9 stars of the basic sample,
homogeneously measured $BVRI$ photometry is available from Bessell (1990) ---
these are used to interpolate the solar colours involving $BVRI$ filters. For
12 stars, $UBV$ photometry is taken from Johnson et al.(1966) --- these are
used to infer the solar $U-B$ colour. For 31 stars, $JHK_s$ data have been
collected from the 2MASS survey.  Finally, for 52 stars, Str\"omgren photometry
is available from Nordstr\"om et al (2004) and Tycho photometry from ESA
(1997).

For each colour, the effective temperatures of the stars are fit to a linear
(first order) function of colour and [Fe/H] (we also tried second order fits,
but these were not significant improvements on linear fitting); in practice, it
was convenient to fit $\theta_{\mathrm eff} = 5040/\Teff$ rather than $\Teff$
directly. For example, for $B-V$, we fitted for $(a_1,a_2,a_3)$ in the relation

\be
\theta_{\mathrm eff} = a_0 + a_1 \times (B-V) + a_2 \times {\mathrm [Fe/H]}.
\ee

\noindent
The reason we fit for $\theta_{\mathrm eff}$ as a function of colour and
metallicity, is because the relative errors in colour and [Fe/H] are always
much smaller than the relative error in $\theta_{\mathrm eff}$. Figure
\ref{tefit} shows the result of the fit.

The relations obtained can then be used to solve for the solar colours
(e.g. $(B-V)_\odot$), under the assumption of $\Teff,\odot = 5777$ K and
[Fe/H]$_\odot = 0.0$. Formally, we have converted the relation to the form
(e.g. for $B-V$)

\be 
B-V = b_0 + b_1 \times (5040/\Teff) + b_2 \times {\mathrm [Fe/H]} \label{feq}
\ee

\noindent
in order to make clearer the effect of temperature and metallicity on derived
colours. Examination of column 4 in table \ref{fits}, for example, shows that
the effect of metallicity in the colour transformations decreases toward redder
colour bands, and reverses its sign in the IR, as one would expect.

The parameters for each of the fitted colours $(b_1,b_2,b_3)$ are shown in
Table~\ref{fits}, and the solar colours derived are shown in
Table~\ref{solcol}. We have carefully propagated the errors in fitting in order
to obtain 1-$\sigma$ error estimates on the derived solar colours.  The error
quoted in Table~\ref{solcol} for each colour is the total dispersion of the
fit. In addition to the direct Tycho $B_{\rm T}-V_{\rm T}$ colour shown in
Table~\ref{solcol}, we also give the resulting Johnson colour derived adopting
the full transformation used for the $B-V$ values given in ESA (1997). The
result is closely consistent with the proper Johnson $B-V$ fit.

We estimate the size of the systematic error as follows: the IRFM temperature
scale is now known to be consistent with the stellar interferometry based
temperature scale 
to within 18~K (dispersion 62 K, standard error 20 K; Ram\'\i rez \& 
Mel\'endez 2005a).  This result is consistent with zero offset, but also
with a systematic offset of order 20 K. An offset of order 20 K produces
systematic shifts in our derived colours of less than half the internal error;
had we included this potential systematic error (in quadrature), our error
estimates in Table \ref{solcol} would increase by 10\% or less for all colours.

Interestingly, there are many more stars with accurate effective temperatures
and metallicities for which photometry is available in the IR and Str\"omgren
colours, than in $UBVRI$ --- since the stars are so bright (and we presently
only have about a dozen stars in the sample) we are planning to extend the
$UBVRI$ sample with observations at the 30 cm telescope on La Palma.

When we tried to extend our study to {\it ugriz} photometry in a similar way as
for the other colours, we found that there is almost no overlap between 
Sloan Digital Sky Survey (SDSS)
standard stars and stars with either IRFM $\Teff$ or [Fe/H] measured. Hopefully
this unfortunate situation will soon be remedied, but in the meantime we apply
a secondary method to obtain the solar colours in the SDSS bands. We have used
the observationally based transformations between SDSS and Johnson--Cousins
photometry from Smith et al. (2002), Karaali, Bilir, \& Tun\c{c}el (2005) and
Bilir, Karaali \& Tun\c{c}el (2005). To give an estimate of the present
uncertainty of the transformations, we have added the dispersion between them
to the propagated dispersion from the fit giving the total number shown in
Table~\ref{solcol}. 

\begin{table}
\begin{center}
\caption{Parameters for the fitted relations (see Eqn \ref{feq}) for each
indicated colour. The number of stars in the fits is shown in the last column.}
\begin{tabular}{lrrrr}  
\hline
Colour          &     $b_0~~~~$&     $b_1~~~~$&   $b_2~~~~$   &  $N$  \\
\hline
$U-B$           & $-4.7081$ & $ 5.5943$ & $ 0.5116$  &  12 \\
$B-V$           & $-1.2929$ & $ 2.2179$ & $ 0.1327$  &   9 \\
$V-R$           & $-0.5958$ & $ 1.0884$ & $ 0.0238$  &   9 \\
$R-I$           & $-0.4036$ & $ 0.8432$ & $-0.0170$  &   9 \\
$V-I$           & $-0.7563$ & $ 1.6550$ & $ 0.0035$  &   9 \\
$B_{T}-V_{T}$   & $-1.9142$ & $ 3.0055$ & $ 0.0984$  &  51 \\
$b-y$           & $-0.8152$ & $ 1.3961$ & $ 0.0276$  &  51 \\
$v-y$           & $-2.2597$ & $ 3.7495$ & $ 0.1478$  &  51 \\
$v-b$           & $-1.4939$ & $ 2.4103$ & $ 0.1201$  &  51 \\
$u-v$           & $-2.9690$ & $ 4.5252$ & $ 0.2212$  &  51 \\
$V-J$           & $-2.4416$ & $ 4.1179$ & $-0.0141$  &  30 \\
$V-H$           & $-2.6302$ & $ 4.6303$ & $-0.0481$  &  30 \\
$V-K_s$         & $-3.0362$ & $ 5.2049$ & $-0.0142$  &  30 \\
\hline
\end{tabular}
\label{fits}
\end{center}
\end{table}

\begin{table}
\begin{center}
\caption{Colours of the Sun, and dispersions of the fit, computed from the
fits to Sun-like stars with accurate colours, effective temperatures and
metallicities. We have assumed $\Teff = 5777$ K and [Fe/H] $ = 0$ for the Sun.}
\begin{tabular}{lc}  
\hline
Johnson/Cousins         & $                  $   \\
$(U-B)_\odot$           & $  0.173 \pm 0.064 $   \\
$(B-V)_\odot$           & $  0.642 \pm 0.016 $   \\
$(V-R)_\odot$           & $  0.354 \pm 0.010 $   \\
$(R-I)_\odot$           & $  0.332 \pm 0.008 $   \\
$(V-I)_\odot$           & $  0.688 \pm 0.014 $   \\
\hline
Tycho                   & $                  $   \\
$(B_{T}-V_{T})_\odot$   & $  0.708 \pm 0.030 $   \\
$(B-V)_\odot$           & $  0.636 \pm 0.023 $   \\
\hline
Str\"omgren             & $                  $   \\
$(b-y)_\odot$           & $  0.403 \pm 0.013 $   \\
$(v-y)_\odot$           & $  1.011 \pm 0.035 $   \\
$(v-b)_\odot$           & $  0.609 \pm 0.023 $   \\
$(u-v)_\odot$           & $  0.979 \pm 0.064 $   \\
\hline
2MASS                   & $                  $   \\
$(V-J)_\odot$           & $  1.151 \pm 0.035 $   \\
$(V-H)_\odot$           & $  1.409 \pm 0.035 $   \\
$(V-K_s)_\odot$         & $  1.505 \pm 0.041 $   \\
\hline
SDSS                    & $                  $   \\
$(u-g)_\odot$           & $  1.40 \pm 0.08   $   \\
$(g-r)_\odot$           & $  0.45 \pm 0.02   $   \\
$(r-i)_\odot$           & $  0.12 \pm 0.01   $   \\
$(i-z)_\odot$           & $  0.04 \pm 0.02   $   \\
\hline
\end{tabular}
\label{solcol}
\end{center}
\end{table}

\section{Discussion}

We have derived colour estimates for the Sun from Sun-like stars, taking
advantage of the fact that the adopted stellar effective temperature scale from
the IRFM has been recently confirmed as both accurate and precise from
interferometric measurements of stellar diameters (Ram\'\i rez \& Mel\'endez
2005a). This is the primary reason for confidence in deriving solar colours
from Sun-like stars, but we have also restricted ourselves to the highest
quality metallicity and colour data as well.

While undertaking this study, we found that solar colours had been derived in a
similar manner by Ram\'\i rez \& Mel\'endez (2005b). They derived broad and
medium band colours : Johnson/Cousins, Vilnius, Str\"omgren $b-y$, DDO, 2MASS
and Tycho photometric systems, as part of a study primarily devoted to the
effective temperature scale of F, G and K stars. We believe we have improved on
their results by (1) severely restricting the stellar sample to the highest
quality metallicities; (2) providing error estimates on the quoted colours and
(3) we concentrate on Sun--like stars in the ($\Teff$, [Fe/H]) region closest
to the Sun, rather than interpolating the solar colours from fitting relations
valid for a much wider range in temperatures and metallicities. Combined with
the requirement of highest quality metallicity data, this restricts our sample
to a rather small number of stars (especially in $UBVRI$), which we plan to
extend in the near future. However we remark that our sample is not
significantly smaller than that of Ram\'\i rez \& Mel\'endez, {\it once the
same restrictions in $T_{\rm eff}$ is applied}.  As mentioned in Section 2, the
requirement of good metallicity data reduces their sample by 40\% (from 115 to
67 stars), and their subsamples with available BVRI photometry would count
11--37 stars vs.\ our 9--12 stars in Table~\ref{fits}. The lower number of
stars in our sample is compensated for by the increased accuracy and
homogeneity in the metallicities, the smaller number of parameters and the
decreased dispersion in the fits for this restricted temperature range (see
below). Notice also that, for $B-V$, our direct determination based on just 9
objects is in excellent agreement with the indirect determination from Tycho
$B_{\rm T}-V_{\rm T}$ colours, which is based on 51 objects.

Actually, all our colours were found to be, within our error estimates, in
excellent agreement with Ram\'\i rez \& Mel\'endez, with the possible exception
of $B-V$. For this colour, we find $(B-V)_\odot = 0.642 \pm 0.016$, while
Ram\'\i rez \& Mel\'endez find $(B-V)_\odot = 0.619$ (no error given). We have
attempted to reconstruct the size of their error, and believe it to be about
twice our own error, or about 0.03 mag. This is based on their estimate of a
scatter in the (10 parameter) fit of effective temperature to colour and
metallicity of 88 K for main sequence stars (their table 2), whereas our own (3
parameter) fit produces a scatter of 43~K. Our reconstruction of the size of 
their error appears to be correct (Ram\'\i rez 2005, priv.\ comm.), hence
the two estimates of $(B-V)_\odot$ agree within
the errors. Sekiguchi \& Fukugita (2000) have also studied the $B-V$
colour--temperature relation, using a very similar method, and deriving
$(B-V)_\odot = 0.626 \pm 0.018$. Our solar colours are also in good agreement
with Cayrel de Strobel (1996): $(B-V)_\odot = 0.642 \pm 0.004$, $(b-y)_\odot =
0.404 \pm 0.005$; Gray (1995): $(B-V)_\odot = 0.648 \pm 0.006$, $(R-I)_\odot =
0.338 \pm 0.002$; Taylor (1998): $(R-I)_\odot = 0.335 \pm 0.002$. 
Another method to determine the colours of the Sun is by applying
synthetic photometry to the observed, absolute flux calibrated solar spectrum; 
this results in $(U-B)_\odot = 0.13-0.14$ and $(B-V)_\odot = 0.63-0.65$ 
(Colina, Bohlin \& Castelli 1996; Bessel et~al.\ 1998), again
in good agreement with our determinations.
Finally, also  the predictions of the most updated theoretical stellar 
atmosphere models: $(B-V)_{\odot} \sim 0.64-0.65$ (Casagrande et~al.\ 2005, 
in prep.) are well compatible with our results.

In summary, we have provided an estimate of the colours of the Sun in a large
variety of photometric systems, by comparison to Sun--like stars in the same
temperature and metallicity range as the Sun.  We made use of the most accurate
and internally homogeneous data available for colours and metallicity, and
took advantage of the recently firmly-established direct and IRFM temperature
scale. We also give a careful estimate of the present uncertainties. Within the
errors, our solar colours are in excellent agreement with the analogous recent
work by Ram\'\i rez \& Mel\'endez (2005b), with the possible exception of $B-V$
which remains somewhat controversial. While in the past the discrepancy was
imputed to the temperature scale (Sekiguchi \& Fukugita 2000), this is now no
longer the issue. The main difference seems to be between studies deriving
$(B-V)_{\odot} \sim 0.62$ from fitting $T_{\rm eff}$--colour relations from a 
wide
range of temperatures and metallicities (Sekiguchi \& Fukugita 2000; Ram\'\i
rez \& Mel\'endez 2005b), and on the other side studies focussing on solar
analogues pre--selected spectroscopically or within a narrower range of
solar--like temperatures (Cayrel de Strobel 1996; Gray 1995; Soubiran \& Triaud
2004; present work) which favour $(B-V)_{\odot} \sim 0.64$. Increased accuracy
in selecting good quality, homogeneous data has also some effect on our
result. Our $(B-V)_{\odot}=0.642$ is in close agreement with the colours of the
closest solar twin known so far (HR 6060) and of the new candidate HIP 78399.
Finally, once errors
are properly taken into account, we remark that there is no crucial discrepancy
between our result and the bluer colour of Sekiguchi \& Fukugita, Ram\'\i rez
\& Mel\'endez. The main limit presently is the small number of stars with good
quality metallicity data, having suitable $UBVRI$ photometry. We plan to
improve on this aspect with new observations in the near future.

\section*{Acknowledgments}
We thank Luca Casagrande for valuable advice on the IRFM and on the
theoretical solar colours; and Ivan Ram\'\i rez for useful correspondence.
This research was supported by the Academy of Finland (grants nr.~206055 and
208792). This publication makes use of data products from the Two Micron All
Sky Survey, which is a joint project of the University of Massachusetts and the
Infrared Processing and Analysis Center/California Institute of Technology,
funded by the National Aeronautics and Space Administration and the National
Science Foundation.

\end{document}